\documentclass[a4paper,fleqn,usenatbib]{mnras}
\pdfoutput=1

\usepackage[T1]{fontenc}
\usepackage{ae,aecompl}

\usepackage{graphicx}	
\usepackage{amsmath}	
\usepackage{amssymb}	

\usepackage{mathptmx}
\usepackage{txfonts}

\newcommand{\Msun}{\mathrm{M_{\sun}}}
\title[Disk and star-formation properties in X-ray groups
  \& clusters]{Comparing galaxy disk
and star-formation properties in X-ray bright and faint groups and clusters}

\author[I.D. Roberts et al.]{
Ian D. Roberts,\thanks{E-mail: roberid@mcmaster.ca}
Laura C. Parker, Ananthan Karunakaran
\\
Department of Physics and Astronomy, McMaster University, Hamilton ON
L8S 4M1, Canada
}

\date{Accepted XXX. Received YYY; in original form ZZZ}

\pubyear{2015}

\begin{document}
\label{firstpage}
\pagerange{\pageref{firstpage}--\pageref{lastpage}}
\maketitle

\begin{abstract}
Galaxy morphologies and star-formation
rates depend on environment.  Galaxies in
underdense regions are generally star-forming and disky whereas
galaxies in overdense regions tend to be early-type and not actively
forming stars.  The mechanism(s) responsible for star-formation
quenching and morphological transformation remain unclear,
although many processes have been proposed.  We study the dependence
of star-formation and morphology on X-ray luminosity for galaxies in
Sloan Digital Sky Survey Data Release 7 (SDSS-DR7) groups and
clusters.  While controlling for stellar and halo mass
dependencies, we find that galaxies in X-ray strong groups and
clusters have preferentially low star-forming and disk
fractions -- with the differences being
strongest at low stellar masses.  The trends that we observe do not
change when considering only galaxies found within or outside of the
X-ray radius of the host group.  When considering central
and satellite galaxies separately we find that this dependence on
X-ray luminosity is
only present for satellites, and we show that our
  results are consistent with ``galaxy stangulation'' as a mechanism
  for quenching these satellites.  We investigate the dynamics of the
groups and clusters in the sample, and find that
the velocity distributions of
galaxies beyond the virial radius in low X-ray luminosity
halos tend
to be less Gaussian in
nature than the rest of the data set.  This may be indicative of low
X-ray luminosity groups and clusters having
enhanced populations of star-forming
and disk galaxies as a result of recent accretion.
\end{abstract}




\section{Introduction}
\label{sec:introduction}
Numerous studies have shown a strong environmental dependence on the
star-forming and
morphological properties of galaxies \citep[e.g.][]{oemler1978,
  dressler1980, postman1984, dressler1999, blanton2005b, wetzel2012}.
Low density
regimes tend to be dominated by star-forming, late-type galaxies
whereas high density areas, such as galaxy clusters, tend to be
primarily populated by quiescent, early-type galaxies.  Within
individual clusters, galaxy morphologies tend to distribute as a
function of local density (or equivalently cluster-centric radius),
with high fractions of late-type galaxies being found at large radii
and the regions near the cluster core being dominated by early-types
\citep[e.g.][]{dressler1980, postman1984, postman2005}.  This
effect has become known as the morphology-density relation.
While galaxies tend to distribute based on their star-forming and
morphological properties, the mechanism(s) responsible for the quenching of
star-formation and morphological transformations in galaxies are not
well constrained -- although many
have been proposed.  Both mergers and impulsive galaxy-galaxy
interactions (``harassment'') \citep[e.g.][]{moore1996} can induce
star-burst events in galaxies leading to rapid consumption of gas
reserves and star-formation quenching.  Within the virial radius of a
group or cluster the stripping of gas from galaxies becomes
efficient.  Both the stripping of hot halo gas (``strangulation'')
\citep[e.g.][]{kawata2008} and cold gas stripping due to a dense
intra-cluster medium (``ram-pressure'') \citep[e.g.][]{gunn1972} can
quench star-formation.  As well, tidal interactions can affect
gas reservoirs by transporting
gas from the galactic halo outwards which subsequently allows it to
more easily be stripped from the galaxy \citep{chung2007}.
\par
On top of these environmental quenching mechanisms, previous authors
have found that secular processes, which depend on galaxy mass,
appear to play a significant role in star-formation quenching
\citep{balogh2004, muzzin2012}.  The emergent picture for star
-formation quenching
appears to be some combination of environmental quenching mechanisms
and internal, secular processes.  In particular, \citet{peng2010}
suggests that in the low redshift Universe, environmental quenching is
dominant for
galaxies with $M_\star \la 10^{10.5}\,\Msun$, whereas for galaxies
with $M_\star \ga 10^{10.5}\,\Msun$ mass quenching plays the more
important role.
\par
While environmental and mass quenching within individual halos are
seemingly strong effects, it is important to realize that groups and
clusters are not isolated structures.  In particular, galaxies can be
pre-quenched in group halos prior to infall into a
larger cluster.  This ``pre-processing'' suggests that many galaxies
may already be quenched upon cluster
infall.  Simulations have shown that between $\sim 25$ and $45$ per
cent of infalling cluster galaxies may have been pre-processed \citep{mcgee2009,
  delucia2012}.  Observationally, \citet{hou2014} find that $\sim 25$
per cent of the infall population reside in subhalos for massive
clusters ($M_H \ga 10^{14.5}\,\Msun$).  This pre-quenching of galaxies
in groups could potentially be driven by galaxy interactions and
mergers which are favoured in the group regime as a result of lower
relative velocities between member galaxies \citep{barnes1985, brough2006}.
\par
An important method for studying the quenching mechanisms in groups
and clusters is to study the dependence of the star-formation and
morphological properties of galaxies on the conditions of their host
halo (e.g.\ halo mass, X-ray luminosity, etc.).  In particular, if
quenching mechanisms depend on the density of the intra-group/cluster medium
(IGM/ICM) -- for example, ram-pressure stipping of cold gas -- then one
would expect to see galaxy populations which are preferentially
passive in halos with high X-ray
luminosities.  Such correlations have been looked for in previous
studies, primarily within cluster environments.  
\par
In particular, \citet{ellingson2001} finds no
positive correlation between the fraction of old galaxies and X-ray
gas density.  \citet{balogh2002} conclude that the level of
star-formation found in their ``low-$L_X$'' sample is consistent with
the
levels seen in their CNOC1 sample consisting of higher mass clusters.
\citet{fairley2002} and \citet{ wake2005} both study the fractions of blue
galaxies at intermediate redshifts and find no discernible trend
between blue fraction and X-ray luminosity.  Using multivariate
regression \citet{popesso2007} find that cluster star-formation
depends on cluster richness but find no additional dependence on X-ray
luminosity.  In addition, they find no significant correlation between
star-forming fraction and any global cluster property ($M_{200}$,
$\sigma_v$, $N_{\mathrm{gal}}$, and $L_X$).  \citet{lopes2014} find no
dependence of blue fraction on X-ray
luminosity and the only slight dependence they find between disk fraction and
X-ray luminosity is within the central and most dense regions.
\par
Conversely, \citet{balogh2002b} find that galaxies in their ``low-$L_X$'' sample
have preferentially high disk fractions compared to galaxies in their
``high-$L_X$'' sample.  \citet{postman2005} find that the
bulge-dominated fraction for galaxies in high X-ray luminosity
clusters is higher than for those in low X-ray luminosity clusters.
In contrast with their star-formation results, \citet{popesso2007} do
find a significant anti-correlation
between blue fraction and X-ray luminosity.
Finally, \citet{urquhart2010} find an anti-correlation between blue
fraction and X-ray temperature for galaxies in intermediate redshift
clusters.
\par
In this paper we re-visit the dependence of galaxy star-formation and
morphological properties on the X-ray luminosity of the host halo.
Specifically, as a result of the large SDSS X-ray sample presented in
\citet{wang2014}, we are able to control for stellar mass, halo
mass, and radial dependencies through fine-binning of the data set.
This allows us
to more directly investigate the effect of X-ray
luminosity on galaxies in different environments.
\par
The results of this study are presented as follows.  In
\S~\ref{sec:data} we briefly describe the SDSS group catalogues
utilized in this work, as well as the star-formation and
morphology catalogues which we match to the group data set.  In
\S~\ref{sec:results} we present the primary results 
of this paper, specifically, the differences between star-forming and
morphological trends in environments with different X-ray luminosities.  In
\S~\ref{sec:discussion} we provide a discussion of the results
presented in this paper.  Finally, in
\S~\ref{sec:summary} we provide a summary of the key results and make
concluding statements.
\par
In this paper we assume a $\Lambda$ cold dark matter cosmology with
$\Omega_M=0.3$, $\Omega_\Lambda=0.7$, and
$H_0=70\,\mathrm{km}\,\mathrm{s^{-1}}\,\mathrm{Mpc^{-1}}$.


\section{Data}
\label{sec:data}

\subsection{Yang group catalogue}

This work relies heavily on the group catalog of
\citet{yang2007}.  The Yang group catalogue is constructed by applying
the iterative halo-based group finder of \citet{yang2005, yang2007} to the New
York University Value-Added Galaxy Catalogue (NYU-VAGC;
\citealt{blanton2005}), which is based on the Sloan Digital Sky Survey
Data Release 7 (SDSS-DR7; \citealt{abazajian2009}).  The Yang group
catalogue has a wide range of halo masses, spanning from
$\sim10^{12}\,\Msun$ to $\sim10^{15}\,\Msun$.  The catalogue
contains both objects which would be classified as groups
($10^{12} \la M_H \la 10^{14}\,\Msun$) and as
clusters ($M_H \ga 10^{14}\,\Msun$), however for brevity we will
refer to all systems as groups regardless of mass. 
\par
Groups are initially
populated using the traditional friends-of-friends (FOF) algorithm
\citep[e.g.][]{huchra1982}, as well
as assigning galaxies not yet linked to FOF groups as the centres of
potential groups.  Next, the characteristic luminosity, $L_{19.5}$,
defined as the combined luminosity of all group members with
$^{0.1}M_r-5\log h \le -19.5$, is calculated for each group.  Using the
value of $L_{19.5}$ along with an assumption for the group
mass-to-light ratio, $M_H/L_{19.5}$, a tentative halo mass is assigned
on a group-by-group basis.  The tentative halo mass is used to
calculate a virial radius and velocity dispersion for each group,
which are then used to add or remove galaxies from the system.
Galaxies are assigned to groups under the
assumption that the distribution of galaxies in phase space follows
that of dark matter particles -- the distribution of which is assumed
to follow a spherical NFW profile \citep{navarro1997}. This process is
iterated until the group memberships no longer change.
\par
Final halo masses given in the Yang group catalogue
are determined using the ranking of the characteristic stellar mass,
$M_{\star,\mathrm{grp}}$, and assuming a relationship between
$M_H$ and $M_{\star,\mathrm{grp}}$ \citep{yang2007}.
$M_{\star,\mathrm{grp}}$ is defined by Yang et al. as

\begin{equation}
  M_{\star,\mathrm{grp}} = \frac{1}{g(L_{19.5},\,L_{\mathrm{lim}})}\sum_i
  \frac{M_{\star,i}}{C_i},
\end{equation}

\noindent
where $M_{\star,i}$ is the stellar mass of the $i$th member
galaxy, $C_i$ is the completeness of the survey at the position of
that galaxy, and $g(L_{19.5},\,L_{\mathrm{lim}})$ is a correction
factor which accounts for galaxies missed due to the magnitude limit
of the survey.  The statistical error in $M_H$ is on the order of
$0.3\,\mathrm{dex}$ and mostly independent of halo mass \citep{yang2007}.

\subsection{SDSS X-ray catalogue}

To study the X-ray properties of the group sample, we utilize the SDSS
X-ray catalogue of \citet{wang2014}.
Wang et al. use ROSAT All Sky
Survey (RASS) X-ray images in conjuction with optical groups
identified from SDSS-DR7 \citep{yang2007} to estimate X-ray
luminosities around $\sim65\,000$ spectroscopic groups..
\par
To identify X-ray luminosities for individual groups, the algorithm of
\citet{shen2008} is employed.  Beginning from an optical group, the
most massive galaxies (MMGs) of that group are identified -- up to
four MMGs are kept.  The RASS fields in which the MMGs reside are then
identified, and an X-ray source catalogue is generated in the
$0.5-2.0\,\mathrm{keV}$ band.  The maximum X-ray emission density
point is used to identify the X-ray centre of the group, and any X-ray
sources not associated with the group (e.g. point source quasars or
stellar objects cross-matched from RASS and SDSS-DR7), within one virial radius,
are masked out.  Values for the X-ray background, centred on the X-ray
centre, are determined and subtracted off and the X-ray luminosity, $L_X$, is
calculated by integrating the source count profile within the X-ray
radius.
\par
Determining X-ray luminosities in this manner is succeptible to
``source confusion''.  Due to
projection it is possible for more than one group to contribute to the
X-ray emission within the X-ray
radius, leading to an overestimation of the X-ray luminosity for a
given group.  To account for this effect Wang et al. calculate the
``expected'' average X-ray flux, $F_{X,i}$, for each group using the average
$L_X - M_H$ relation taken from \citet{mantz2010}.  They then
calculate the sum of the expected fluxes from each group for
multi-group systems and
determine the contribution fraction, $f_{\mathrm{mult},i}$, for each
group defined as,

\begin{equation}
  f_{\mathrm{mult},i} = F_{X,i}/\Sigma_i F_{X,i}.
\end{equation}

\noindent
The contribution factor will approximate the fraction of the observed
X-ray luminosity intrinsic to the individual group in question,
therefore applying this fraction to each group will act to debias
the measured X-ray luminosity from source confusion contamination.
\par
Within the Wang catalogue 817 groups have $S/N > 3$, compared to the
total of 34522 groups with positive detections (positive count rates
after background subtraction) and $S/N > 0$.  We run our analysis for
groups with $S/N >3$ as well as groups with $S/N > 0$ and find that
our choice of signal-to-noise cut does not change the trends that we
observe, therefore we focus on the total sample
($S/N > 0$) to ensure a sample size which is large enough to finely
bin the data in various properties simultaneously.

\subsection{Final data set}

\begin{figure}
  \centering
  \includegraphics[width=\columnwidth]{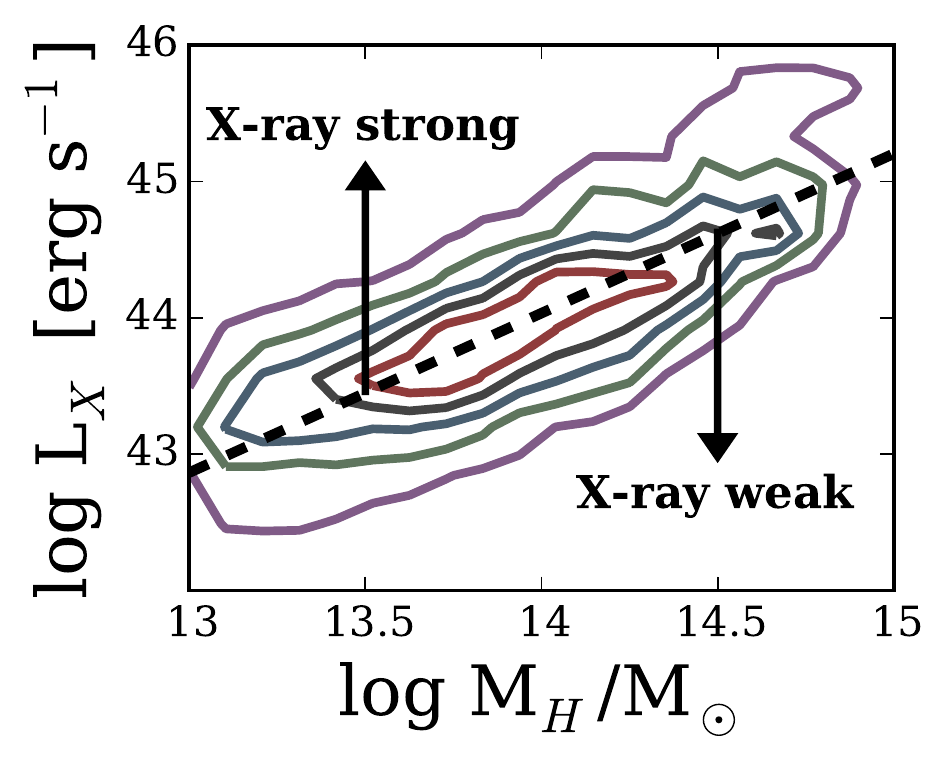}
  \caption{Density contours for log X-ray luminosity versus log halo
    mass.  Dashed line
  corresponds to the linear least-squares best-fit relationship.}
  \label{fig:mh_lx}
\end{figure}

\begin{figure}
  \centering
  \includegraphics[width=\columnwidth]{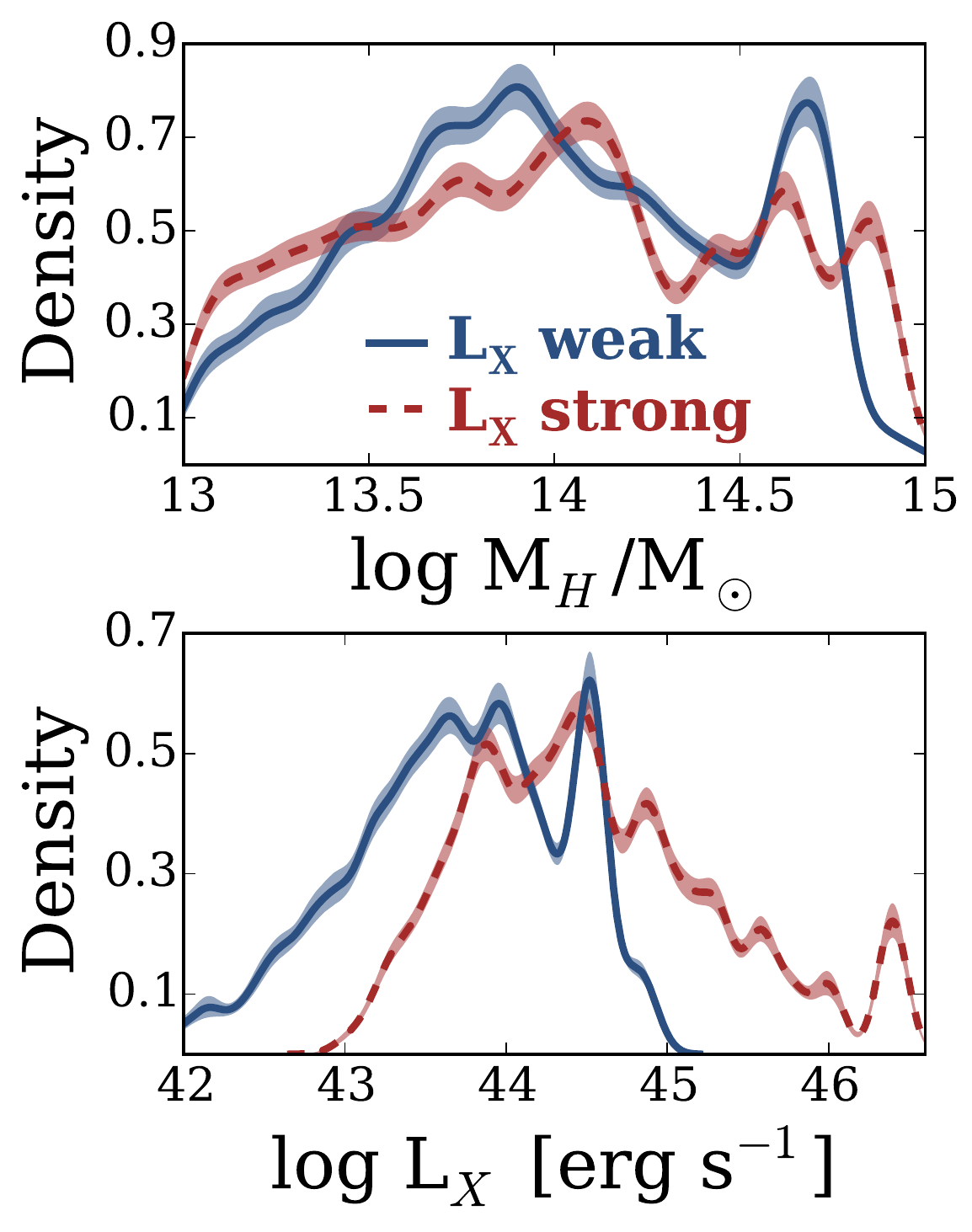}
  \caption{Smoothed distributions for halo mass and X-ray
    luminosity within the sample.  Distributions
    are shown for both the X-ray strong (red, dashed) and the X-ray weak
    (blue, solid) samples.
    Shaded regions correspond to $2\sigma$ confidence intervals
    obtained from random bootstrap resampling.}
  \label{fig:data_smooth}
\end{figure}

\begin{figure*}
  \centering
  \includegraphics[width=\textwidth]{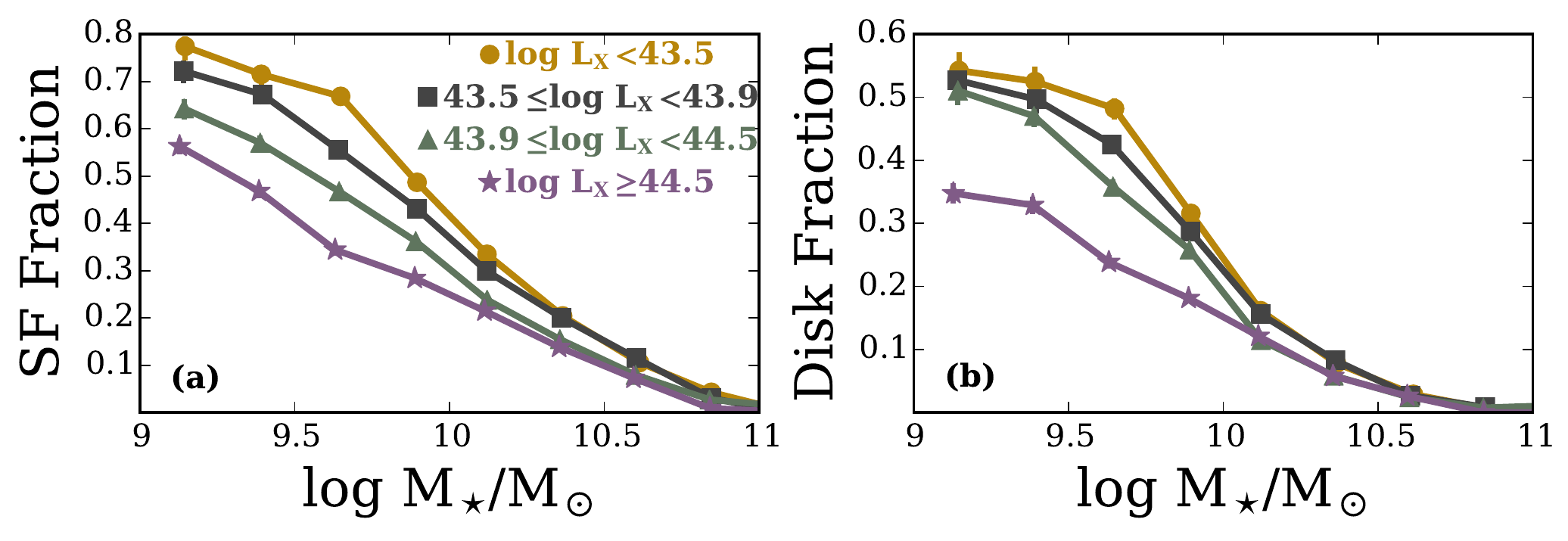}
  \caption{Left: star-forming fraction versus stellar mass for the four
    X-ray luminosity quartiles of the data sample.  Right:
    disk fraction versus stellar mass for the four
    X-ray luminosity quartiles of the sample.  Error bars
    correspond to 1$\sigma$ Bayesian binomial confidence
    intervals given in \citet{cameron2011}.}
  \label{fig:df_pf_lx}
\end{figure*}

To obtain the final data set, we match the Wang
SDSS X-ray catalogue to the Yang SDSS group catalogue, giving us
both optical and X-ray group properties for the sample.  To
obtain individual galaxy properties we further match the data set to
various public SDSS catalogues, as follows.
\par
We utilize stellar masses given
in the NYU-VAGC, which are computed following the methodology of
\citet{blanton2007}.
\par
To obtain star-formation rates (SFRs) and specific star formation
rates ($\mathrm{SSFR} = \mathrm{SFR}/M_{\star}$) we match the catalogue of 
\citet{brinchmann2004} to the sample.  SFRs given by
Brinchmann et al. are determined using emission line fluxes whenever
possible, however in the case of no clear emission lines or
contamination from active galactic nuclei, SFRs are determined using
the strength of the 4000 {\AA} break ($D_n4000$) \citep{brinchmann2004}.
\par
We obtain galaxy morphologies from the catalogue of
\citet{simard2011}.  Simard et al. perform two-dimensional bulge + disk
decompositions for over one million galaxies from the Legacy area of
the SDSS-DR7, using three different fitting models: a pure S\'{e}rsic
model, a bulge + disk model with de Vaucouleurs ($n_b = 4$) bulge, and
a bulge + disk model with a free $n_b$.  To distinguish between disky
and elliptical galaxies we utilize the galaxy S\'{e}rsic index, $n_g$,
from the pure S\'{e}rsic decomposition.  We also use the $V_{\mathrm{max}}$
weights given by Simard et al. to correct for the incompleteness of
our sample.
\par
We calculate group-centric distances for each galaxy using the
redshift of the group and the angular seperation between the galaxy and the
luminosity-weighted centre of its host group.  We normalize all of the
galaxy radii by the virial radius of the host group, $R_{180}$, which
we calculate as in \citet{yang2007}

\begin{equation}
  R_{180} =
  1.26\,h^{-1}\,\mathrm{Mpc}\left(\frac{M_H}{10^{14}\,h^{-1}\mathrm{M_\odot}}\right)^{1/3}(1
    + z_g)^{-1},
\end{equation}

\noindent
where $z_g$ is the redshift of the group center.

\begin{figure*}
  \centering
  \includegraphics[width=\textwidth]{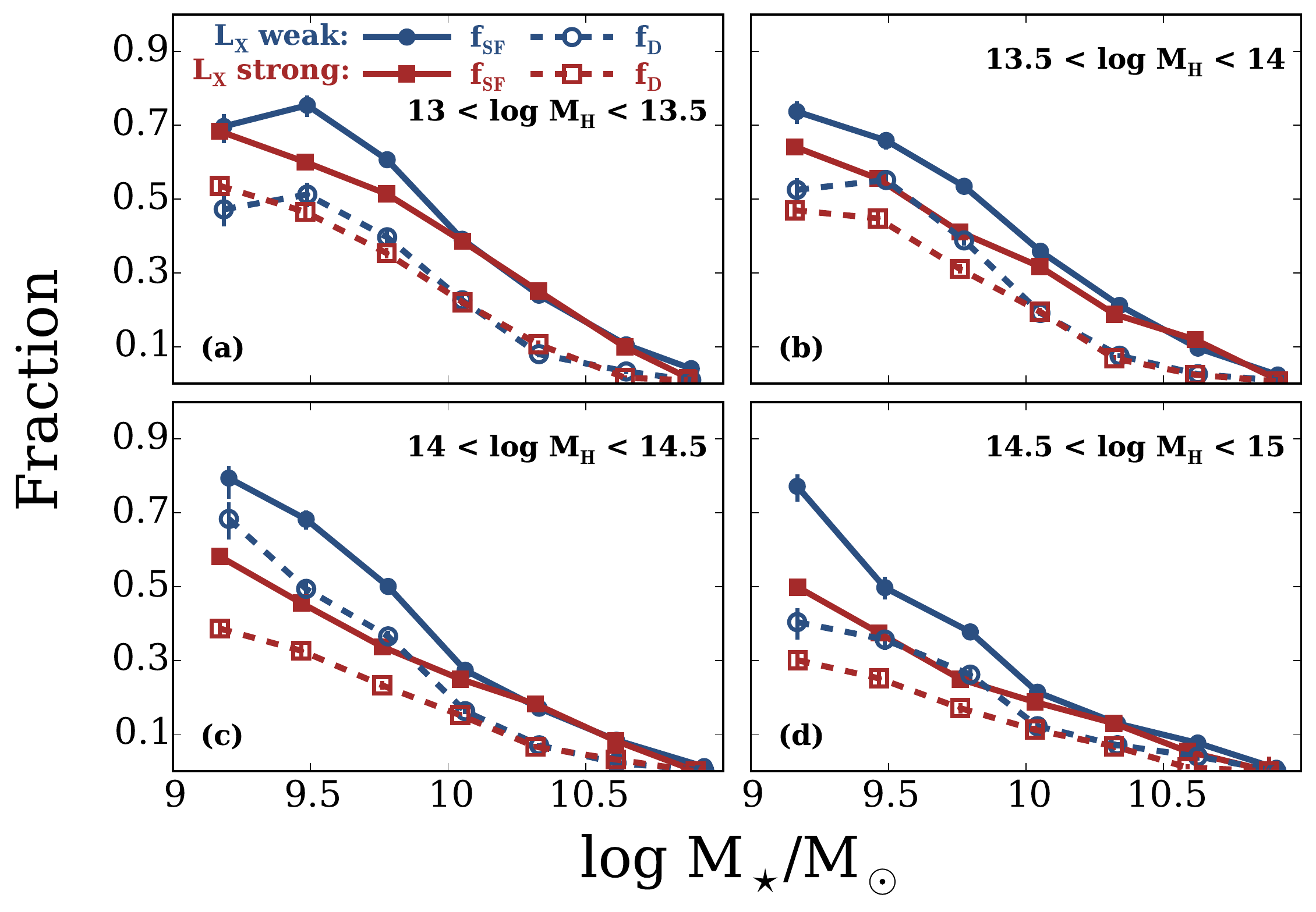}
  \caption{Star-forming (solid lines) and disk (dashed lines) fractions
    versus stellar mass, for different halo mass bins and the XRW
    (blue) and XRS (red) samples.  Error bars
    correspond to 1$\sigma$ Bayesian binomial confidence
    intervals given in \citet{cameron2011}.}
  \label{fig:f_lx_mh}
\end{figure*}

The final data set includes groups with halo masses ranging between
$10^{13} - 10^{15}\,\Msun$, and galaxies with stellar masses ranging from
$10^9 - 10^{11.3}\,\Msun$.  Group X-ray luminosities in the data set are
between $10^{39.6} - 10^{46.4}\,\mathrm{erg}\,\mathrm{s^{-1}}$, with a median value
of $10^{43.9}\,\mathrm{erg}\,\mathrm{s^{-1}}$, and are strongly
correlated with halo
mass (see Fig.~\ref{fig:mh_lx}).  We do not make an explicit radial
cut, however over 99 per cent of member galaxies fall within 1.5
virial radii.  Our final sample contains 3902 low redshift ($z<0.1$)
groups housing a collective 41173 galaxies.  The
    catalogue of
    \citet{wang2014} contains $\sim 35000$ groups, the fact that the
    final sample in this work is significantly smaller is for two
    reasons.  Firstly,
    we restrict our sample to redshifts smaller than 0.1 which reduces
the number of groups from $\sim 35000$ at $z<0.2$ to $\sim 18000$ at
$z<0.1$.  The second important cut is that we require $10^{13} < M_H < 10^{15}\,\Msun$,
a number of groups in the Wang et al. catalogue have halo masses, $M_H
< 10^{13}\,\Msun$ (where halo masses have been obtained from the catalogue of
\citealt{yang2007}), this cut reduces the remaining number of groups
from $\sim 18000$ to $\sim 3900$.  It should be noted that the
majority of the $M_H < 10^{13}\,\Msun$ groups removed from the data
set are groups with very low membership.
\par
To determine the effect of X-ray luminosity on star-formation and
morphology we consider two X-ray luminosity samples for the majority
of our analysis, which we refer to as the X-ray weak (XRW) and X-ray
strong (XRS) samples. Similar to Wang et al., we define the XRS sample
to consist of
all galaxies found above the best-fit $\log M_H - \log L_X$ line (see
Fig.~\ref{fig:mh_lx}), and correspondingly the XRW sample consists of
all galaxies found below the $\log M_H - \log L_X$ trend line.  This leads to an
approximately equal number of galaxies within the XRW and XRS
samples.  We also performed
our analysis with a cut between the two X-ray samples
at the median X-ray luminosity of the data
set, as well as
defining the two samples using the first and fourth quartiles, however
these alternative definitions of the two X-ray samples do not 
change the trends that we observe.
\par
Smoothed distributions for halo mass and X-ray luminosity are shown in
Fig.~\ref{fig:data_smooth} for both X-ray luminosity samples.  Density
distributions are calculated using the
\texttt{density \{stats\}} function in the statistical computing language
\texttt{R} \citep{r2013}\footnote{http://www.R-project.org/} using a
Gaussian kernel.
\par
We study the dependence of star-formation rates and morphology on
stellar mass by binning the data by stellar mass and calculating the
disk and star-forming fractions for each bin.  Binning by stellar mass
is important to account for the systematic dependence of star-formation and
morphology on stellar mass
\citep[e.g.][]{brinchmann2004, whitaker2012}.  Additionally, as the relative
balance between environmental and mass quenching is not well
understood, it is important to investigate the effects of environment
at a given stellar mass.  
\par
We define the star-forming fraction,
$f_{SF}$, as the fraction of galaxies in
each bin with $\log \mathrm{SSFR} > -11$.  \citet{wetzel2012} show that at
low redshift the division between the red sequence and the blue cloud
is found at $\log \mathrm{SSFR} \simeq -11$ across a wide range of halo masses.
For each stellar mass bin the star-forming fraction is given by

\begin{equation}
  f_{SF} =
  \frac{V_{\mathrm{max}}\;\text{weighted}\;\#\;\text{galaxies}\;\text{with}\;\log SSFR>-11}{V_{\mathrm{max}}\;\text{weighted}\;\text{total}\;\#\;\text{galaxies}}.
\end{equation}

\noindent
Similarly we define the disk fraction, $f_D$, as the fraction of
galaxies in each bin with S\'{e}rsic index, $n < 1.5$.  For each
stellar mass bin this is given by

\begin{equation}
  f_D =
  \frac{V_{\mathrm{max}}\;\text{weighted}\;\#\;\text{galaxies}\;\text{with}\;n<-1.5}{V_{\mathrm{max}}\;\text{weighted}\;\text{total}\;\#\;\text{galaxies}}.
\end{equation}

\noindent
We also ran our analysis using a dividing cut at S\'{e}rsic indices of
$n=1.0$ and
$n=2.0$ to define a disk galaxy, however using
these alternative definitions for a disk galaxy does not alter the
trends that we observe.


\section{Results}
\label{sec:results}

\begin{figure*}
  \centering
  \includegraphics[width=\textwidth]{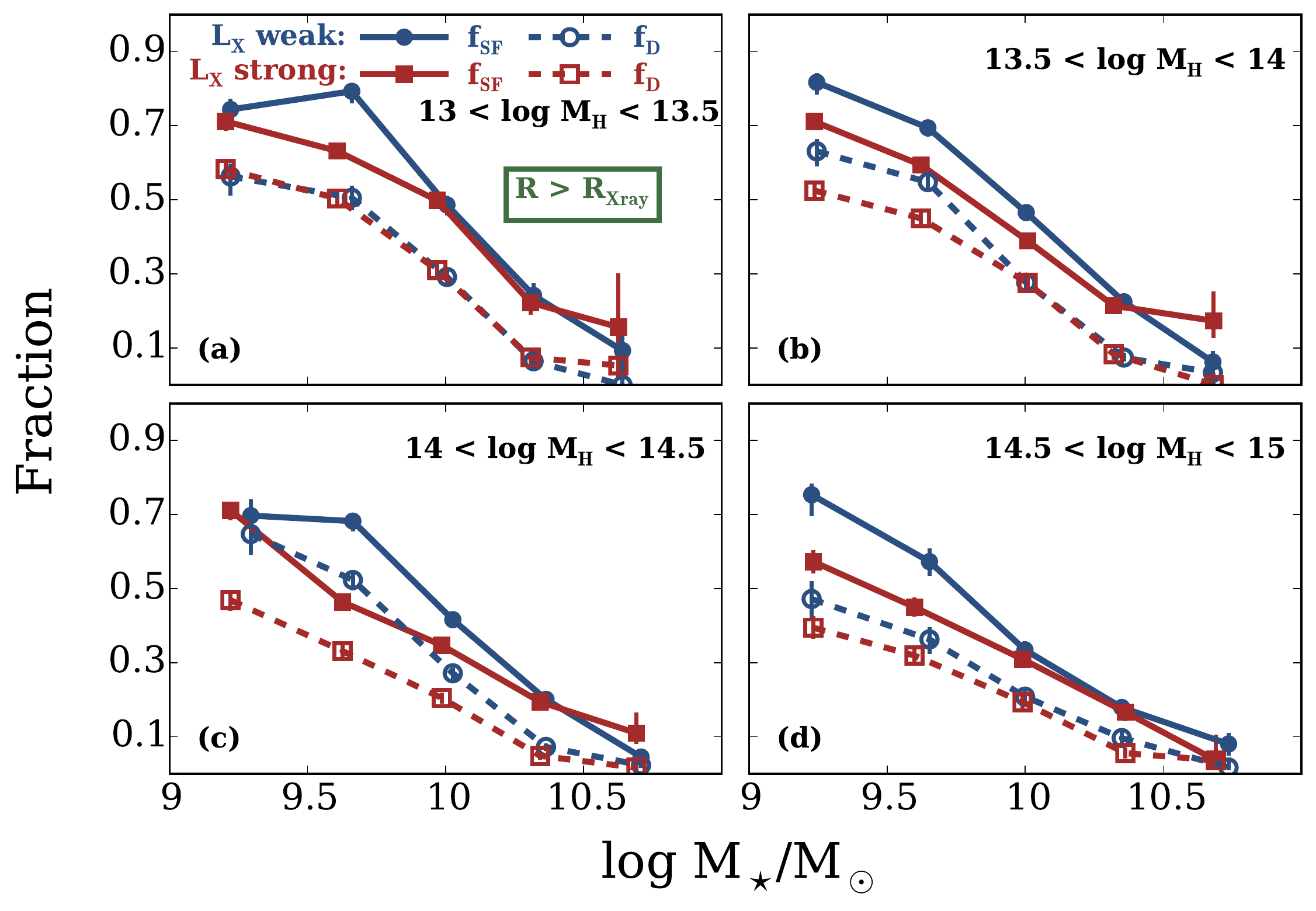}
  \caption{Star-forming (solid lines) and disk (dashed lines)
    fractions versus stellar mass, for galaxies outside of their host
    X-ray radius and for different halo mass bins and
    the two $L_X$ samples.  Error bars
    correspond to 1$\sigma$ Bayesian binomial confidence
    intervals given in \citet{cameron2011}.}
  \label{fig:f_lx_mh_rxh}
\end{figure*}

\begin{figure*}
  \centering
  \includegraphics[width=\textwidth]{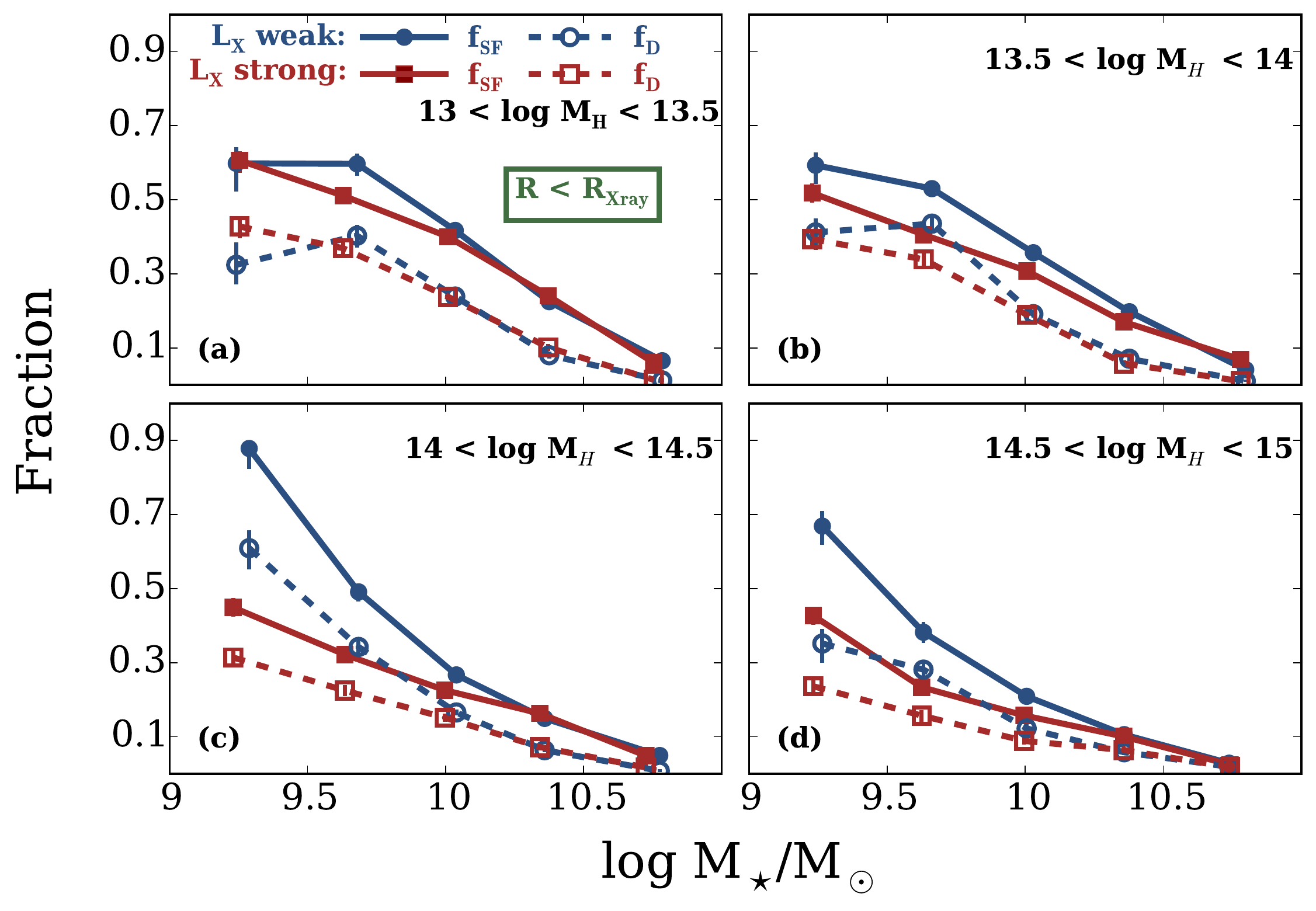}
  \caption{Same as Fig.~\ref{fig:f_lx_mh_rxh} for galaxies inside of
    their host X-ray radius.}
  \label{fig:f_lx_mh_rxl}
\end{figure*}

\begin{figure*}
  \centering
  \includegraphics[width=\textwidth]{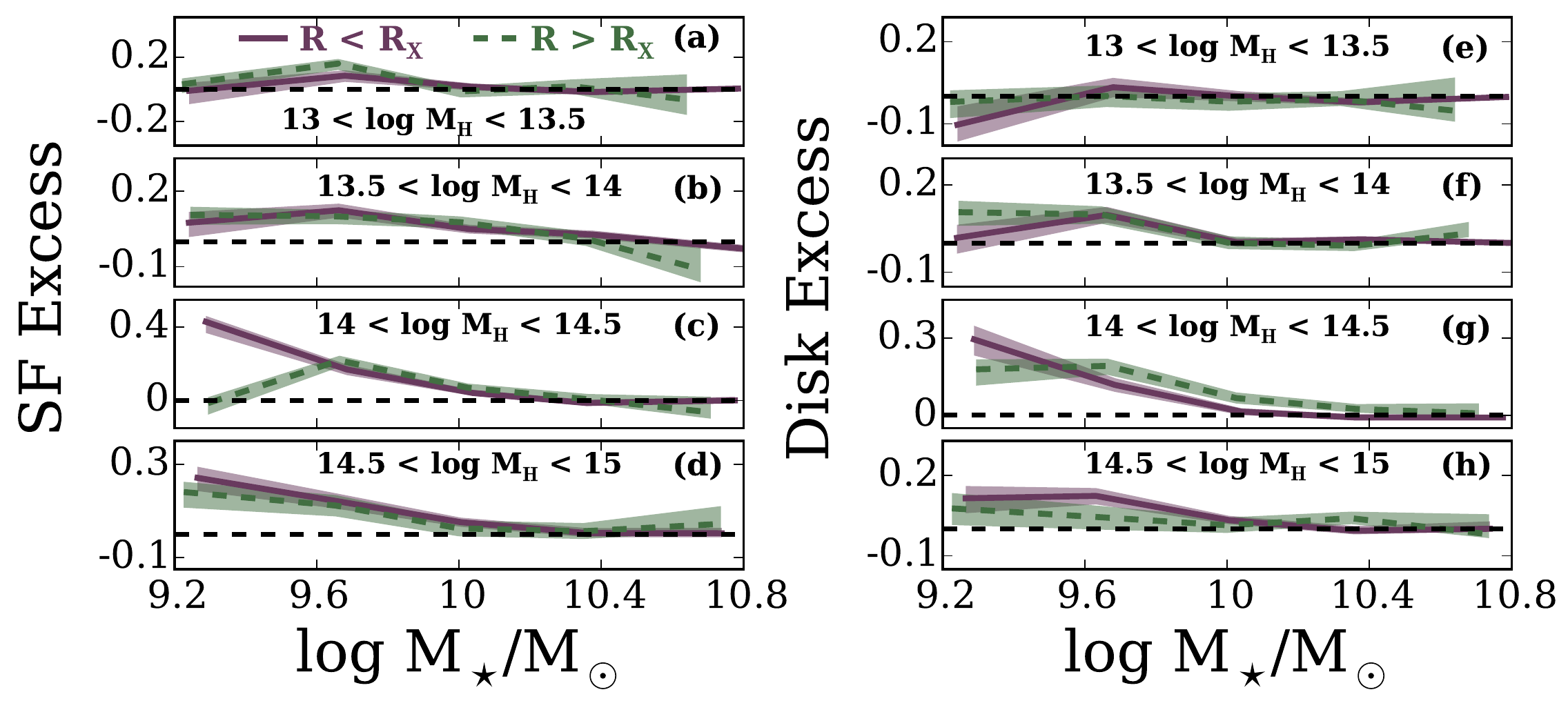}
  \caption{SF and disk excess versus stellar mass for both galaxies
    within (purple, solid) and
    outside (green, dashed) of the X-ray radius.  Panels a-d show SF excess
    for four halo mass bins and panels e-h show disk excess for
    four halo mass bins.  Shaded regions represent
  $1\sigma$ confidence intervals.}
  \label{fig:lxDiff_sf_d}
\end{figure*}

\subsection{Star-forming and morphology trends in strong and weak $L_X$
  samples}
\label{sec:bin_trend}

To investigate the effect of X-ray luminosity on galaxy properties, in
Fig.~\ref{fig:df_pf_lx} we show star-forming and disk fractions, as a function
of stellar mass, for subsamples corresponding to the four X-ray
luminosity quartiles of the data set.
Examination of Fig.~\ref{fig:df_pf_lx}a and \ref{fig:df_pf_lx}b show
that star-forming and disk fractions follow a consistent marching order
with respect to X-ray luminosity.  The disk and star-forming fractions
decrease as X-ray luminosity increases.
\par
We note that the results in Fig.~\ref{fig:df_pf_lx} consider all halo
masses in the sample, however it has been found that galaxy morphology and
star-formation depend on local density and halo mass
\citep{dressler1980, balogh2004, wetzel2012, lackner2013}
(however also see: \citealt{delucia2012, hoyle2012, hou2013}).  As shown in
Fig.~\ref{fig:mh_lx} the data show a strong correlation between X-ray
luminosity and halo mass, therefore we must determine if differences
shown in Fig.~\ref{fig:df_pf_lx} are simply a result of galaxies in
higher $L_X$ environments being housed
in preferentially high-mass halos.
\par
To control for any potential halo mass effect, we further bin the data into
narrow halo mass bins and re-examine the dependence of galaxy
properties on X-ray
luminosity, considering now the XRW and XRS
samples from Fig.~\ref{fig:mh_lx}.  Fig.~\ref{fig:f_lx_mh} shows
star-forming (solid) and disk (dashed) fractions as a function of
stellar mass for four different halo
mass bins -- ranging from $10^{13} - 10^{15}\,\Msun$ with bin widths
of $0.5\,\mathrm{dex}$.  Data are binned according to stellar mass and
markers are plotted at the median bin values.  For each halo mass bin
we show star-forming and disk
fractions from the X-ray strong and X-ray weak samples.
\par
For both star-forming and disk fractions we continue to see a residual trend
with X-ray luminosity, even after controlling for any halo mass
dependence: star-forming and disk fractions are systematically higher
in the XRW sample.  We see the strongest trends in the intermediate
and high mass
halos.  The difference between the
strong (red) and the weak (blue) X-ray luminosity samples is clearest at low stellar
mass, and in all halos the two samples converge at
moderate to high stellar mass.

\subsection{Radial dependence of star-forming and morphology trends}

Within host groups X-ray emission is concentrated at relatively small
group-centric radii, with X-ray emission generally extending out to
half a virial radius \citep{wang2014}.  If the
trends we are observing are a result of increased gas density, we
would expect to see enhanced trends (i.e.\ a larger difference between
the XRS and XRW
samples) at small group-centric radii and suppressed trends at
large radii.  To test this we further divide the data into subsets
corresponding to those galaxies that lie within the X-ray emission
radius (using the X-ray radius, $R_{Xray}$, given in \citealt{wang2014}) and
those galaxies that lie outside of the X-ray radius.  We again plot
star-forming/disk fraction versus stellar mass, in narrow halo mass bins,
for the large and small radius subsamples.  The results of this analysis
are shown in Fig.~\ref{fig:f_lx_mh_rxh} \& \ref{fig:f_lx_mh_rxl}, where
the two figures correspond to disk fraction and star-forming fraction trends
for the large and small radius subsamples, respectively. 
\par
Examination of Fig.~\ref{fig:f_lx_mh_rxh} \& \ref{fig:f_lx_mh_rxl}
shows that for both galaxies found within their host halo's X-ray
radius, and those found outside, we still see an increase in
star-forming and disk fractions in the XRW sample -- as before this
effect is strongest in the intermediate to high mass halos and at low
stellar mass. Also the disk and star-forming
fractions tend to be higher at large radii, which is consistent with
the morphology-density relation.
\par
To further investigate if the increase
in star-forming and disk fractions in the XRW sample compared to the
XRS sample -- which we will
refer to as the ``SF excess'' and the ``disk excess'' -- depends on
whether you consider galaxies within or outside the X-ray radius, we
show SF and disk excess versus stellar mass in
Fig.~\ref{fig:lxDiff_sf_d}. We quantitatively define
SF and disk excess as

\begin{align}
  & \mathrm{SF}\;\mathrm{excess} = f_{SF}(XRW) -
  f_{SF}(XRS) \label{eq:sf_excess} \\
  & \mathrm{Disk}\;\mathrm{excess} = f_{D}(XRW) -
  f_{D}(XRS) \label{eq:d_excess}, 
\end{align}

\noindent
where $f_{SF}(XRW)$ and $f_{SF}(XRS)$ are the star-forming fractions in
the XRW and XRS samples respectively, and analogously for $f_{D}(XRW)$
and $f_{D}(XRS)$.
\par
We find no radial dependence for SF and disk excess as the two radial
subsamples in Fig.~\ref{fig:lxDiff_sf_d} show
overlap for all halo and stellar masses.  With the exception in
Fig~\ref{fig:lxDiff_sf_d}c where the SF
excess, for low-mass galaxies, is stronger for galaxies within
the X-ray radius.


\section{Discussion}
\label{sec:discussion}

We find that star-forming and disk fractions are systematically lower in
the XRS sample than galaxies in XRW
environments.  This trend persists even upon controlling for any halo
mass dependence, however the observed difference between the XRS
and the XRW sample is not enhanced when considering
only those galaxies within the X-ray radius of the host halo.
\par
There are two major observed effects which have been found to impact
the distributions of
early-type and late-type galaxies within cluster environments.  The so
called ``Butcher-Oemler'' (BO) effect is the observational trend that
the blue fraction of cluster galaxies are positively correlated with redshift
\citep[e.g.][]{butcher1984, ellingson2001, loh2008, urquhart2010}.  However, it
should be noted that there is still debate when it comes to the
physical nature of the BO effect (for example, see:
\citealt{andreon1999, andreon2004, andreon2006}).  Since we are only
considering low-redshift $(z < 0.1)$ galaxies the BO effect should be
negligible.

\begin{figure}
  \centering
  \includegraphics[width=\columnwidth]{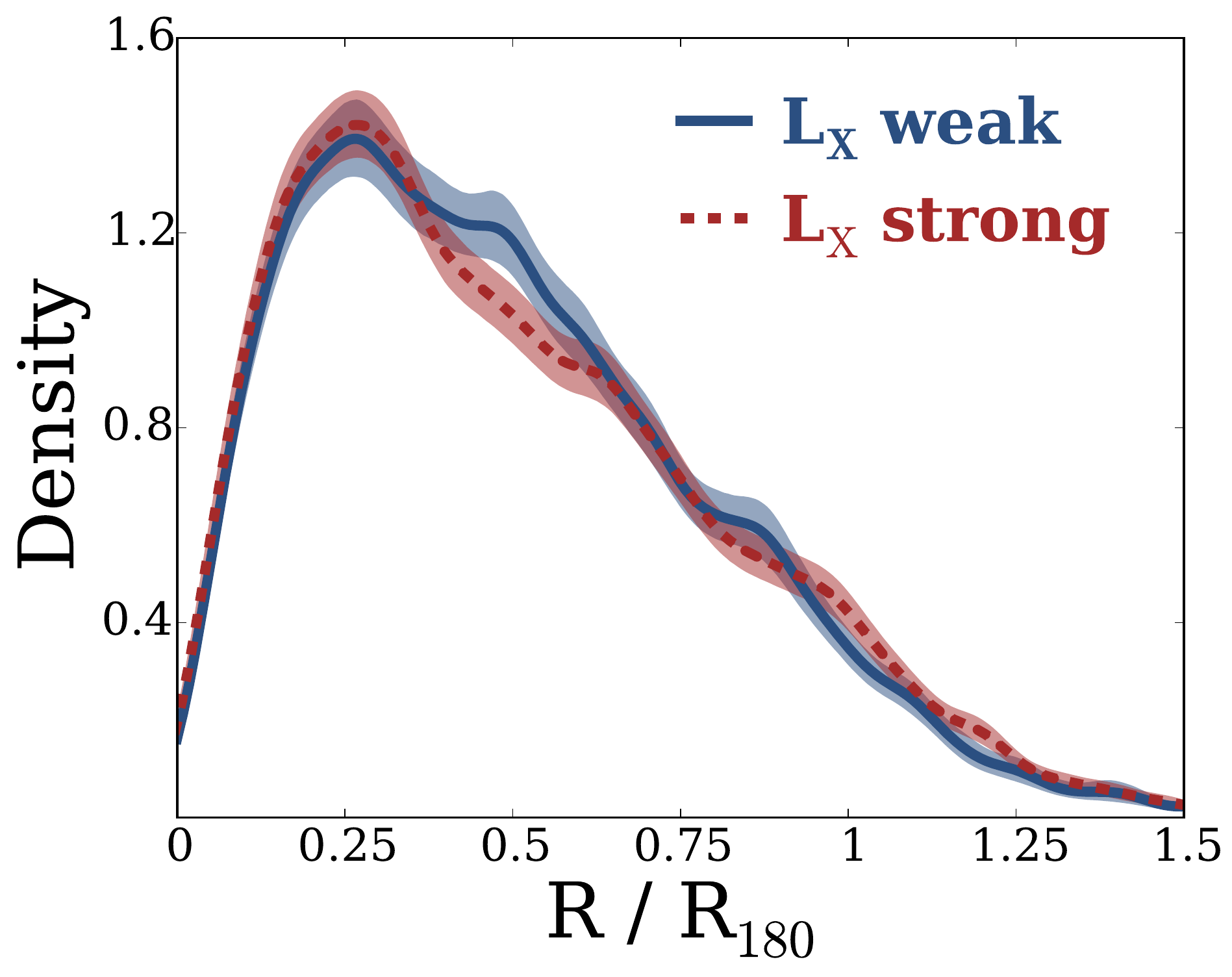}
  \caption{Smoothed radial distributions of galaxies in the XRW (blue,
    solid) and XRS (red, dashed) samples.
    Shaded regions correspond to $2\sigma$ confidence intervals
    obtained from random bootstrap resampling.}
  \label{fig:r_hist}
\end{figure}

The second major effect is the previously mentioned morphology-density
relationship.  In order to determine if the morphology-density
relation is affecting the trends we observe, we must check if there are
significant differences
in the radial distributions of the XRS and the XRW
samples.  For
instance, if the XRW sample is found at systematically high
group-centric radii compared to the XRS sample, then the
morphology-density relation could explain why we find systematically larger
star-forming and disk fractions in the XRW sample.  In
Fig.~\ref{fig:r_hist} we plot the smoothed radial
distributions for both the XRS and the XRW samples.  We
see no systematic difference between the two distributions, in fact
they are nearly indistinguishable from one another.  We 
conclude that the two X-ray samples have radial distributions
which do not differ substantially from one another, and therefore any
observed differences
between the XRS and XRW samples are not being driven by
differing radial distributions.

\subsection{AGN contamination}

\begin{figure}
  \centering
  \includegraphics[width=\columnwidth]{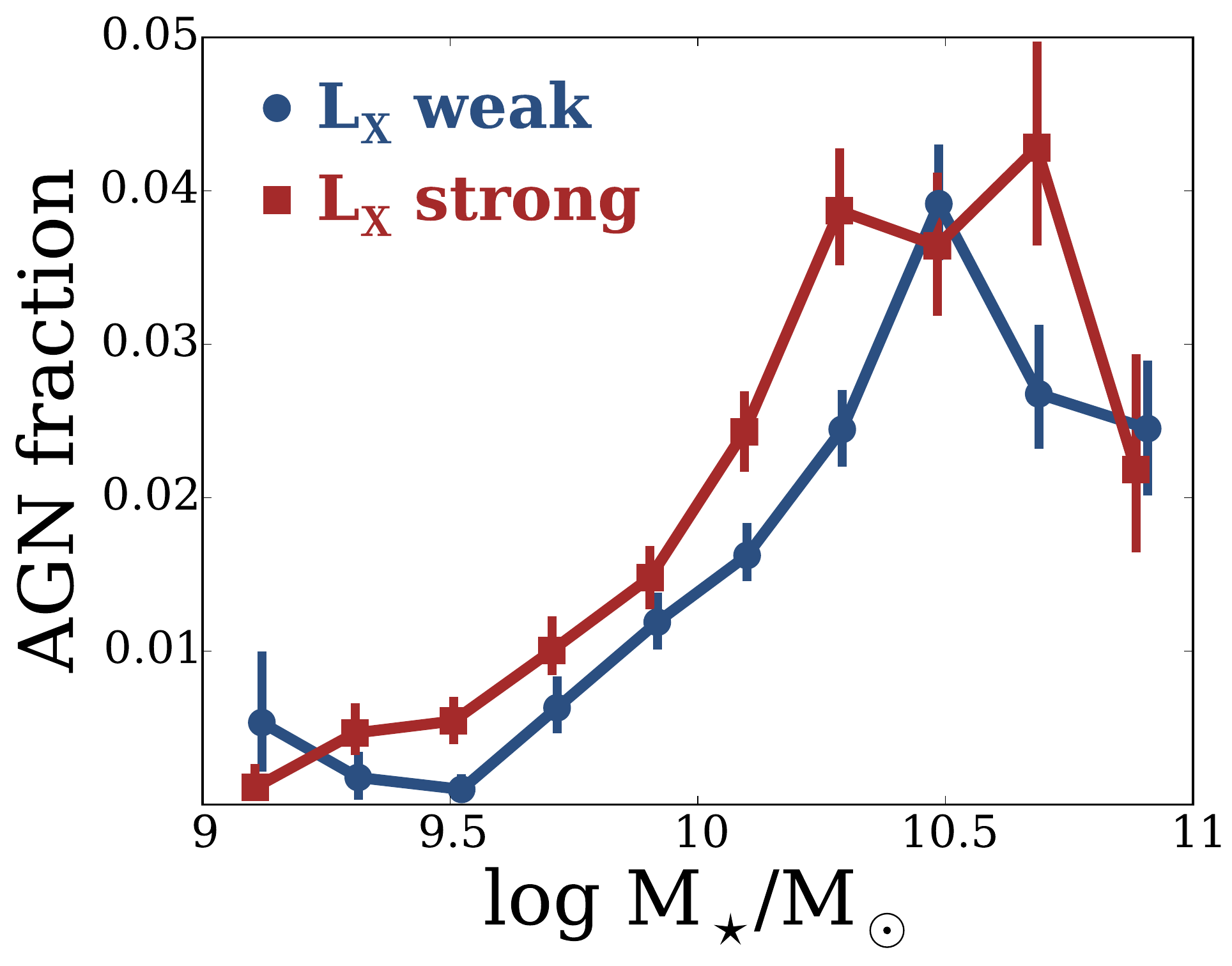}
  \caption{BPT identified AGN fraction versus stellar mass for our XRW and XRS
    samples.  Error bars correspond to $1\sigma$
    Bayesian binomial confidence intervals given in \citet{cameron2011}.}
  \label{fig:AGNFrac}
\end{figure}

When considering X-ray properties of galaxy groups it is important to
ensure that the observed X-ray emission is due to the hot IGM and
not due to contamination from AGN or other X-ray sources.  In
\citet{wang2014} bright point sources, such as stars and quasars, are
masked out, however it is still important to ensure that
our results are not being contaminated by galaxies housing non point
source AGN.
\par
In Fig.~\ref{fig:AGNFrac} we plot AGN fraction versus stellar mass for
the XRW and XRS samples.  We use AGN classified by
\citet{kauffmann2003}, which are identified using the location of
galaxies on the BPT diagram \citep{baldwin1981}.  It should be noted
that \citet{trouille2010} show that between 20 and 50 per cent
(depending on the dividing line between AGN and star-formings galaxies used) of
X-ray identified AGN fail to be classified as AGN on the BPT diagram.
\par
We see that the AGN fraction tends to be
larger within the XRS sample, however at all stellar masses the number of AGN
galaxies is a modest fraction (less than five per cent) of the total
sample, for both XRS and XRW galaxies.  Most relevant is the fact that
at low stellar mass the AGN
fraction is consistently below one per cent, for both the XRW and XRS
samples, and that the trends that we observe with X-ray luminosity are
exclusively seen at low stellar mass (e.g. Fig~\ref{fig:f_lx_mh}).  We
examined disk and
star-forming fractions for a subsample of the data with galaxies
identified as AGN removed and
found that removing AGN galaxies from the sample does not change the
observed trends.  Furthermore, we examined trends
    after removing all groups that
    house galaxies identified as AGN and again found no change in
    the observed trends.  Therefore we conclude that AGN are
  not a significant
contributor to the observed trends in star-forming and disk fractions.

\subsection{Implications for star-formation quenching}

\begin{figure}
  \centering
  \includegraphics[width=\columnwidth]{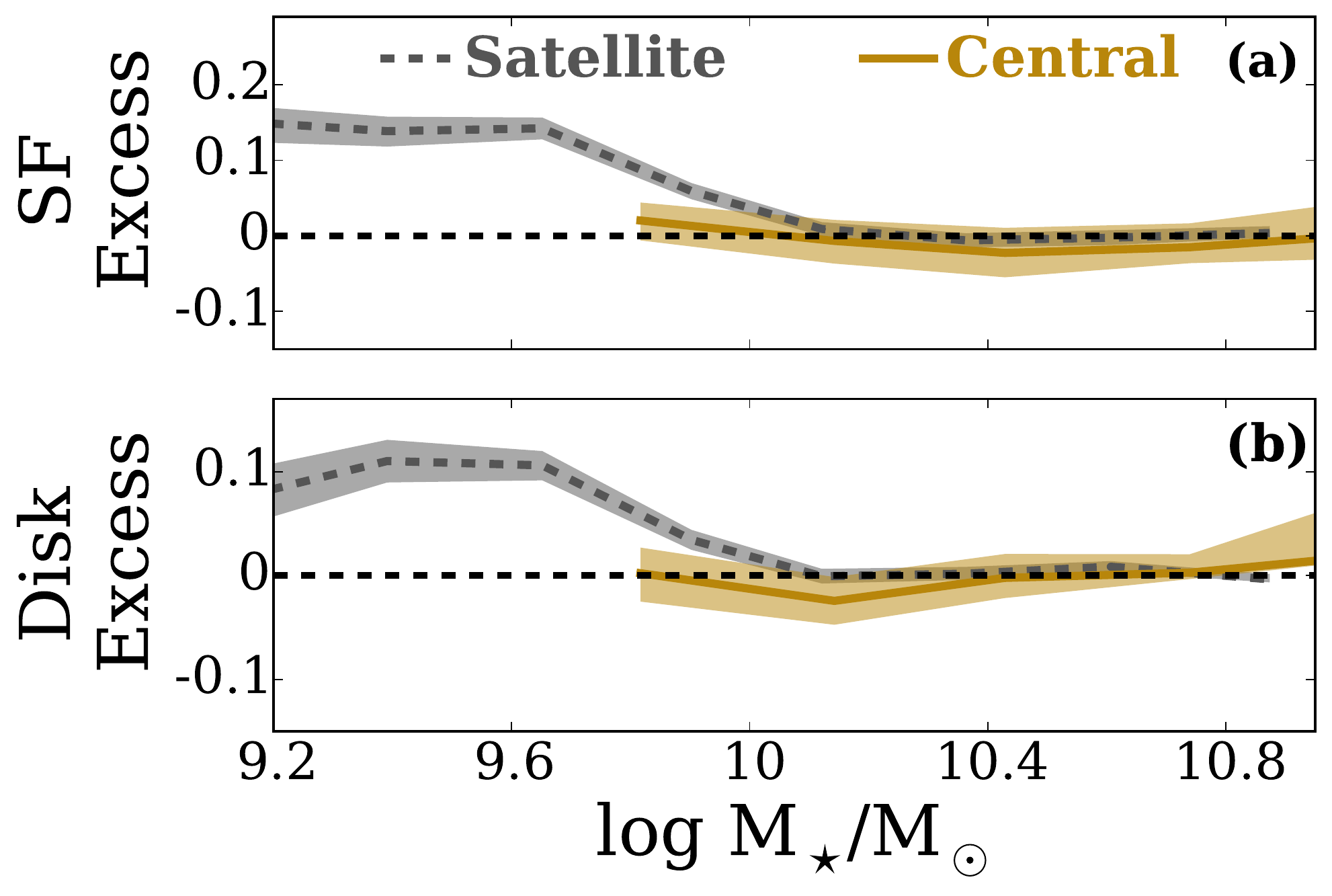}
  \caption{SF and disk excess versus stellar mass for both centrals
    (gold, solid) and satellites (gray, dashed).  Shaded regions
    correspond to $1\sigma$ confidence
  intervals.}
  \label{fig:lxDiff_cs}
\end{figure}

The relative importance of various galaxy quenching mechanisms is an
important, open question.  Galaxy populations in groups
can be classified as either ``central'' (located at the centre
of the group dark matter halo) or ``satellite''
galaxies.  These two populations are
expected to evolve differently \citep[e.g.][]{vandenbosch2008}, and
therefore when
attempting to elucidate information on the quenching of galaxies it is
important to consider centrals and satellites as distinct populations.
In Fig.~\ref{fig:lxDiff_cs} we
plot SF and disk excess (Equations~\ref{eq:sf_excess} \&
\ref{eq:d_excess}) versus stellar mass, considering
separately central and satellite galaxies.  Central galaxies are
defined as the most massive group
galaxies and satellite galaxies are defined as all group galaxies which have
not been classified as centrals.

When considering satellite galaxies in Fig.~\ref{fig:lxDiff_cs}a we find
that galaxies within the XRW sample have consistently larger
star-forming fractions at low stellar mass (SF excess $> 0$), while at
large stellar mass the
XRW and XRS samples are indistinguishable. When considering only
central galaxies we find that there is no
difference between the XRW and XRS samples (SF excess $\approx 0$) when
considering star-forming fraction.  We observe qualitatively similar
trends for disk excess in Fig.~\ref{fig:lxDiff_cs}b.  This implies that
whatever effect
X-ray luminosity has on star-forming and morphlogical properties it
only affects satellite galaxies, central galaxies are insensitive to
the group X-ray properties.  This is not surprising given that central
galaxies are massive, and we see no difference between the XRS and XRW
at large stellar mass.
\par
One interpretation of the differences we observe between the XRW and
XRS samples would be to invoke the ram-pressure stripping of satellite
galaxies.  The rate at which galaxies will lose gas through
ram-pressure stripping will increase in proportion to $L_X$
\citep{fairley2002}.  Therefore, if ram-pressure is an important
mechanism when it comes to the quenching of galaxies, a decrease
in star-forming fraction should be observed with increasing X-ray
luminosity.  It be noted that
although we observe very similar trends for star-forming and disk
fractions, it is not clear whether ram-pressure stripping can
efficiently drive galaxy morphology transformations
from late to early type \citep{christlein2004}.  Prior studies
\citep[e.g.][]{gavazzi2003, kenney2004, muzzin2014} have found
evidence of ram-pressure
stripping. We note as well that other studies \citep[e.g.][]{balogh2002,
  fairley2002, wake2005, lopes2014} have found no clear trend between
star-forming or blue fractions and X-ray luminosity.  At first glance
the results shown in Fig.~\ref{fig:f_lx_mh} are consistent with ram
pressure stripping; at low stellar masses there are lower star-forming
fractions in the XRS sample.  One difference
between the results we observe and previous studies is that we narrowly bin our
data in stellar mass.
Since star-forming and morphological properties depend strongly on
stellar mass, any residual dependence on X-ray luminosity may be lost
without controlling for stellar mass.  In addition our sample size is
significantly larger than most previous studies, so it may be that
trends with X-ray luminosity are subtle enough to be missed without
large statistics.
\par
If the trends we detect are driven by ram-pressure we would expect a
radial dependence of our trends with X-ray
luminosity.  The efficiency of ram-pressure stripping
is proportional to $\rho v^2$ \citep{wake2005,
  popesso2007}, where $\rho$ is the IGM density and $v$ is
the speed of the member galaxies.  Since the IGM density is highest at
small group-centric radii, the efficiency of ram-pressure stripping
should increase towards small radii.  In Fig.~\ref{fig:lxDiff_sf_d} we
showed that the observed SF excess does not strongly depend on radius.
We conclude that this lack of radial dependence is
inconsistent with the ram-pressure stripping scenario.  

\begin{figure}
  \centering
  \includegraphics[width=\columnwidth]{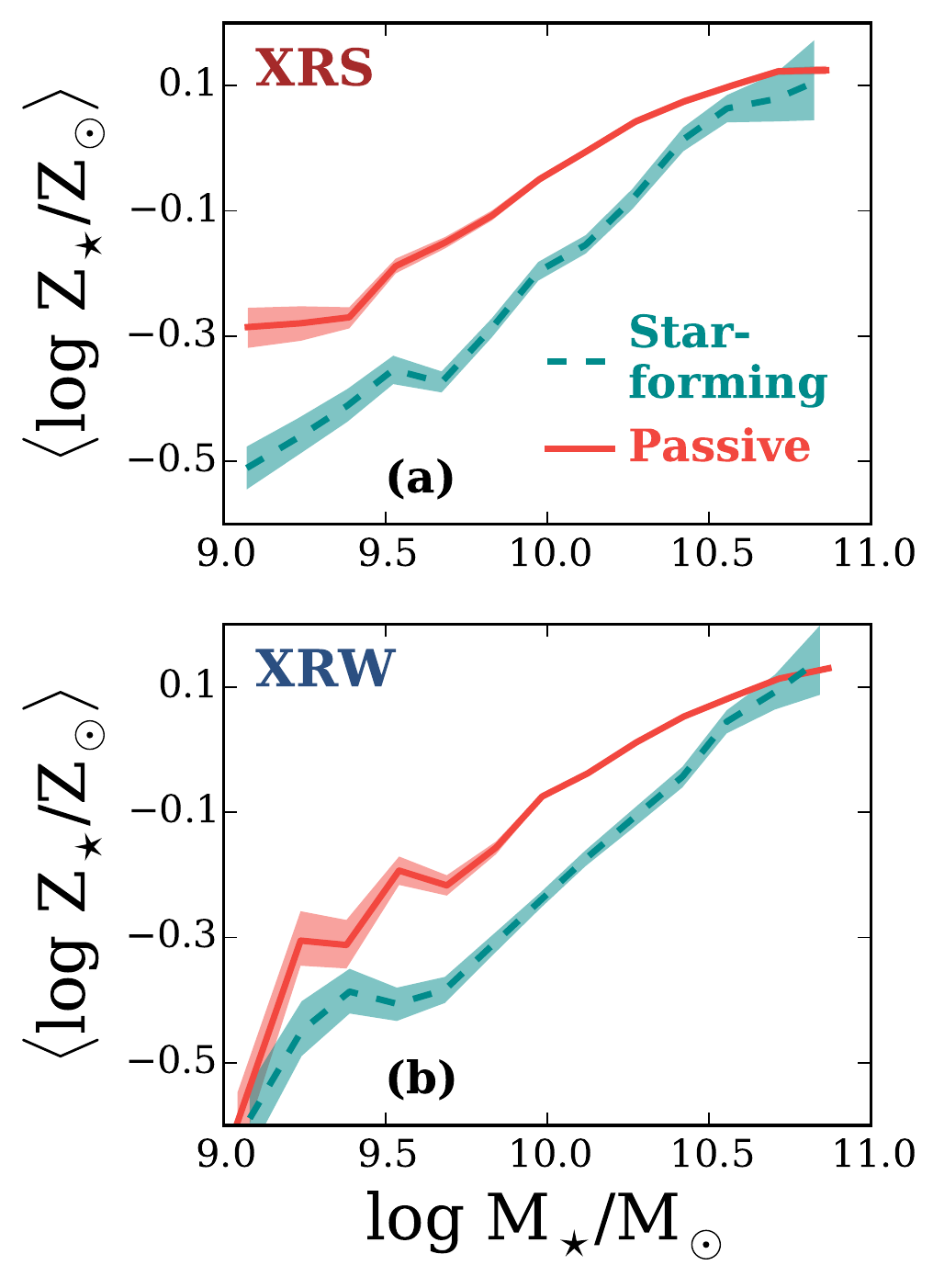}
  \caption{Mean stellar metallicity versus stellar mass for
    star-forming (blue, dashed) and passive (red, solid) galaxies,
    divided by galaxies in
  the XRS (top) and XRW (bottom) samples.  Shaded regions correspond
  to $1\sigma$ confidence intervals obtained from random bootstrap resampling.}
  \label{fig:met_m}
\end{figure}

Another often envoked mechanism for regulating star-formation is
``galaxy strangulation'' \citep{larson1980, balogh2000, kawata2008,
      peng2015}.  Strangulation is a mechanism in
which the replenishment of cold gas onto galaxies is halted, which in
turn leads to galaxy quenching once the galaxy has exhausted its
existing cold gas reservoirs.  The timescales over which a galaxy will
be quenched by strangulation are longer than the quenching times
associated with the direct stripping of cold gas reserves
(ram-pressure).  Recently, \citet{peng2015} have argued that it
is possible to differentiate between strangulation and direct
stripping using metallicity differences between star-forming and
quiescent galaxy populations.  We direct the reader to
\citet{peng2015} for a more complete discussion, however the main idea
is that quenching by strangulation will result in higher
metallicities for passive galaxies compared to star-forming galaxies.
This is a result of star-formation continuing even after the gas
supply has been halted which will increase stellar metallicity until
the cold gas reserves have been exhausted and the galaxy has therefore
been quenched.  This trend in metallicity is not expected from direct
stripping, where star-formation shuts off quickly after the removal of
cold gas.
\par
To investigate the effect of strangulation on the galaxy sample in
this study we follow Peng et al. and calculate mean stellar
metallicity
versus stellar mass considering star-forming and passive galaxies
separately, for galaxies within our XRW sample as well as our XRS
sample.  Metallicities are matched to our sample from the catalogue of
\citet{gallazzi2005}, and mean metallicities are plotted in stellar
mass bins with widths of $0.15\,\mathrm{dex}$.  Not all of the
galaxies within this sample have measured metallicities, therefore for
this aspect of the analysis our XRW and XRS samples are reduced to
10939 (52 per cent of total sample) and 8851 (44 per cent of total
sample) member galaxies, respectively.
\par
In Fig.~\ref{fig:met_m} we see higher stellar
    metallicities for passive galaxies compared to star-formers, which
    we interpret as evidence for strangulation playing a significant role
in star-formation quenching.  Of particular interest for this work is
the behaviour at low stellar mass which is where the dependence of
star-formation and morphology on X-ray luminosity is observed (see
Fig.~\ref{fig:f_lx_mh}).  We see
a somewhat stronger strangulation signal (ie.\ difference between passive and
star-former metallicity) for galaxies in the XRS sample compared to
the XRW sample, at low stellar mass.
\par
In light of this observed difference, it important
    to note that compiling this subsample of
    galaxies with measured metallicities does not affect all stellar
    masses equally.  Specifically, low-mass galaxies are
    preferentially removed from the sample when matching to the
    metallicity catalogue.  In particular, 69 per cent of low-mass
    ($M_\star < 10^{9.5}\,\Msun$) galaxies in the XRS sample do not have
    measured metallicities, whereas in the XRW sample 75 per cent of
    low-mass galaxies do not have measured metallicities.  Not only
    are low-mass galaxies being preferentially lost, but the fraction
    of low-mass galaxies being lost is slightly different between the
    two X-ray samples.  Therefore, although the results in
    Fig.~\ref{fig:met_m} are consistent with strangulation -- and more
specifically, somewhat stronger strangulation at the low-mass end of
the XRS sample -- we suggest that this trend be interpretted with
caution as completeness differences could be playing some role.

\subsection{Group evolutionary/dynamical state}

The dynamical state of galaxy groups is an important evolutionary
indicator and can potentially have an impact on galaxy
properties.  Trends with X-ray luminosity may reflect that the XRW and
XRS samples have different dynamical properties as it is expected that
more evolved groups with relaxed dynamics would be more X-ray luminous
\citep{popesso2007b}.
\par
Theoretically the velocity distribution of galaxies
within a group in dynamical equilibrium should have a characteristic
Gaussian shape.  Groups lacking this Gaussian distribution can
therefore be considered as being unevolved, dynamically young
systems.  To investigate the dependence on the dynamical state of the
groups in our data set we follow the procedure of
\citet{hou2009} and apply the Anderson-Darling normality (ADN) test to
the velocity distributions of the galaxies in the group sample.  The ADN test
is a non-parametric test which compares the cumulative
distribution function (CDF) of the data to the CDF of a normal
distribution to determine the probability (p-value) that the
difference between the distribution of the data and that of a
Gaussian is as large as observed (or larger), under the assumption that the data
is in fact normally distributed.  For our dynamical analysis we use a
subset of the data
consisting of only those groups with eight or more members (31820
galaxies in 1456 groups), in order to
ensure reasonable statistics when applying the ADN test.
To obtain values for the ADN statistic for each
of our groups we employ the
\texttt{ad.test \{nortest\}} function in the statistical computing language
\texttt{R} \citep{r2013} -- large values of the
    ADN statistic are indicative of less Gaussian distributions.
\par
Initially, we examine the dynamical states of galaxies within the XRW
and XRS samples globally (i.e.\ no radial cuts) and we find no
systematic differences between the dynamical states of XRW and XRS
galaxies.  \citet{popesso2007b} study the difference between X-ray underluminous
Abell (AXU) clusters and normal Abell clusters.  They find that while
both AXU and normal Abell clusters show Gaussian velocity
distributions within the virialized region $(R < 1.5\,R_{200})$,
within the exterior regions $(1.5\,R_{200} \le R \le 3.5\,R_{200})$
the AXU clusters show sharply peaked, non-Gaussian velocity
distributions.  The authors interpret these leptokurtic velocity
distributions in the outer cluster regions as evidence that AXU
clusters have experienced recent
accretion/merging.  If the XRW groups have experienced more recent
accretion of galaxies from
the field and smaller groups than the XRS groups, then this could
contribute to the dependence we
observe between star-forming and disk fractions on X-ray luminosity.
Galaxies in underdense regions (the field, low-mass groups) have been
found to be preferentially star-forming with late-type morphologies.
Accordingly, groups experiencing recent accretion may contain more
star-forming, late-type, galaxies when compared to groups which are
dynamically older.
\par
To investigate this possibility we study the dynamical states of
groups in both the XRW and XRS samples, and divide member galaxies
into two radial subsamples: those found in the inner regions $(R <
R_{180})$ of their host group, and those found in the outer regions
$(R \ge R_{180})$ of their host group.  This is similar to the
analysis performed by \citet{popesso2007b}. Instead of making an
arbitrary, discrete cut to define Gaussian and
non-Gaussian groups we treat the AD statistic
values as continuous and compare
the distributions of ADN statistics from the
four subsamples (XRW inner, XRW outer, XRS inner, XRS outer) to
determine whether there are any significant differences in
dynamical states. 
To quantitatively compare the distributions we utilize the two-sample
Anderson Darling (AD2) test.  The AD2 test is similar to the ADN test,
however instead of comparing observed data to the
normal distribution, it compares the CDFs of two data samples to
determine whether they are drawn from the same underlying
distribution.  We apply the AD2 test to the distributions of ADN
statistic values for the XRW and XRS samples to
determine if the dynamical
states vary between the inner and outer regions.  To perform the AD2
test between the subsamples we use the
\texttt{ad.test \{kSamples\}} function in the statistical computing language
\texttt{R} \citep{r2013}.
\par
We find
no evidence $(\text{p-value} = 0.38)$ for different dynamical
states in the inner and outer regions of the XRS sample, however
for the XRW sample we find strong evidence $(\text{p-value} = 3
\times 10^{-7})$ that the dynamical state of galaxies in the outer
region is different from those in the inner region.  When we examine
the distributions of ADN statistics for the four
subsamples we find that
the ADN statistic values for the XRW outer
subsample are systematically higher
than for the other three subsamples.  This suggests that the
velocity distributions for galaxies outside of the virial radius in the XRW
sample are less Gaussian than the rest of the data set.
\par
This result is consistent with \citet{popesso2007b}, who find
non-Gaussian velocity distributions for galaxies in the outer regions
of X-ray underluminous Abell clusters.  This result supports the
notion that the increased number
of star-forming and late-type galaxies we observe in the XRW
sample can potentially be explained by underluminous X-ray groups experiencing
recent accretion of field galaxies and small galaxy groups, as
this recent accretion can give rise to less Gaussian velocity
distributions in the exteriors of these groups.
\par
We do note that it remains difficult to simultaneously explain the
dynamical results together with the fact that we observe no dependence of
SF and disk excess on radius (Fig.~\ref{fig:lxDiff_sf_d}).


\section{Summary \& conclusions}
\label{sec:summary}

We have used a sample of galaxies taken from X-ray
emitting groups and clusters in the SDSS to study the effect of X-ray
luminosity on galaxy star-formation and morphological properties.  Using a data
set spanning a large range in stellar mass ($10^9 -
10^{11.3}\,\mathrm{M_\odot}$), halo mass ($10^{13} -
10^{15}\,\mathrm{M_\odot}$), and X-ray luminosity ($10^{39.6} -
10^{46.4}\,\mathrm{erg}\,\mathrm{s^{-1}}$) we have investigated the
differences between disk and star-forming fractions within different X-ray
environments. The main results of this paper are as follows:

\begin{enumerate}
  \item Star-forming and disk fractions are preferentially lower
    within the X-ray strong sample when
    compared to galaxies within the X-ray weak sample -- this trend
    remains after controlling for any halo mass dependence. 

  \item This difference between the X-ray strong and X-ray weak samples
    is most apparent at intermediate to high halo mass
    and at low stellar mass.

  \item The differences we observe between the X-ray weak and X-ray
    strong samples do not depend on whether we consider galaxies
    inside of, or outside their host halo's X-ray radius.

  \item The enhancement of star-forming and disk fractions we observe
    in the X-ray weak sample is present for satellites but not central
    galaxies, which is not surprising given that the difference
    between X-ray samples is only seen at low stellar mass.
  
  \item Our results are consistent with quenching by
      strangulation, in particular we see a somewhat stronger
      strangulation signal at low stellar mass within the XRS sample.

  \item We find that in the X-ray weak sample the velocity
    distributions of galaxies outside of the virial radius are less
    Gaussian than galaxies within the virial radius.  We find no
    differences between the dynamical states of inner and outer
    galaxies within the X-ray strong sample.
\end{enumerate}

With the large
sample of SDSS X-ray and spectroscopic groups we are able to study
star-forming and disk fractions while simultaneously controlling for stellar
mass, halo mass, and radial depedencies, thereby allowing a robust
analysis of the effects of X-ray luminosity on star formation and
morphology.
\par
We find that galaxies outside the virial radius of X-ray underluminous
groups have dynamics which are less Gaussian than the other groups in
the sample.  This may indicate that recent accretion onto low X-ray
luminosity groups contributes to an excess of star-forming, late-type
galaxies. The fact that the X-ray weak sample
    shows weaker stangulation could simply be due to the lower $L_X$
    environment reducing the efficiency of strangulation, or it could
    be a result of recently accreted galaxies having had less time to
    be quenched by environmental quenching mechanisms like
    stangulation. Naively, one would
expect to observe a corresponding enhancement of star-forming,
late-type galaxies in the exteriors of low X-ray luminosity groups
compared to X-ray strong groups; however this is not observed. The
results presented in this work therefore require a detailed theoretical
treatment to fully explain the trends observed. 


\section*{Acknowledgments}
\label{sec:acknowledgments}

We thank the anonymous referee for their various helpful comments and
suggestions which have improved the paper.  IDR and LCP thank the National
Science and Engineering Research
Council of Canada for funding.  The authors thank
F. Evans for matching together the various SDSS catalogues used in
this research.  We thank X. Yang et al. for
making their
SDSS DR7 group catalogue publicly available, L. Simard et al. for the
publication of their SDSS DR7 morphology catalogue, J. Brinchmann et al. for
publication of their SDSS SFRs, G. Kauffmann et al. for the publication of
SDSS AGN galaxies, the NYU-VAGC
team for the 
publication of their SDSS DR7 catalogue, and A. Gallazzi et al. for making
publicly available their SDSS metallicities.  This research would not have
been possible without access to these public catalogues.
\par
Funding for the SDSS has been provided by the Alfred P. Sloan
Foundation, the Participating Institutions, the National Science
Foundation, the U.S. Department of Energy, the National Aeronautics
and Space Administration, the Japanese Monbukagakusho, the Max Planck
Society, and the Higher Education Funding Council for England. The
SDSS Web Site is http://www.sdss.org/.
\par
The SDSS is managed by the Astrophysical Research Consortium for the
Participating Institutions. The Participating Institutions are the
American Museum of Natural History, Astrophysical Institute Potsdam,
University of Basel, University of Cambridge, Case Western Reserve
University, University of Chicago, Drexel University, Fermilab, the
Institute for Advanced Study, the Japan Participation Group, Johns
Hopkins University, the Joint Institute for Nuclear Astrophysics, the
Kavli Institute for Particle Astrophysics and Cosmology, the Korean
Scientist Group, the Chinese Academy of Sciences (LAMOST), Los Alamos
National Laboratory, the Max-Planck-Institute for Astronomy (MPIA),
the Max-Planck-Institute for Astrophysics (MPA), New Mexico State
University, Ohio State University, University of Pittsburgh,
University of Portsmouth, Princeton University, the United States
Naval Observatory, and the University of Washington.




\bibliographystyle{mnras}
\bibliography{CompleteManuscriptFile_v2} 


\bsp	
\label{lastpage}
\end{document}